\documentstyle[12pt]{article}

\begin{document}

\begin{titlepage}
\setcounter{page}{1}

\author{Waldemar Puszkarz\thanks{
Electronic address: puszkarz@cosm.sc.edu} \\
{\small {\it Department of Physics and Astronomy,} }\\
{\small {\it University of South Carolina,} }\\
{\small {\it Columbia, SC 29208}}}
\title{{\bf Energy Ambiguity in Nonlinear Quantum Mechanics}}
\date{{\small (May 15, 1999)}}
\maketitle

\begin{abstract}
We observe that in nonlinear quantum mechanics, unlike in the linear theory, 
there exists, in general, a difference between the energy functional defined 
within the Lagrangian formulation as an appropriate conserved component of the 
canonical energy-momentum tensor  and the energy functional defined as the 
expectation value of the corresponding nonlinear Hamiltonian operator. 
Some examples of such ambiguity are presented for a particularly simple model 
and some known modifications. However, we point out that there exist a class 
of nonlinear  modifications of the Schr\"{o}dinger equation where this 
difference does not occur, which makes them more consistent 
in a manner similar to that of the linear Schr\"{o}dinger equation. 
It is found that necessary but not sufficient a condition for 
such modifications is the homogeneity of the modified Schr\"{o}dinger 
equation or its underlying Lagrangian density which is assumed to be 
``bilinear'' in the wave function in some rather general sense. Yet, it is 
only for a particular form of this density that the ambiguity in 
question does not arise. A salient feature of this form is the presence
of phase functionals. The present paper thus introduces a new class of 
modifications characterized by this desirable and rare property.

\vskip 0.5cm \noindent

\end{abstract}
\end{titlepage}

%

\section{Introduction}

Nonlinear equations are widespread in physics. Some of them have a
fundamental status, to name Einstein's equations in general relativity or
nonlinear equations of non-Abelian gauge theories. On the other hand, there
is a good deal of nonlinear equations that are a result of some natural
physical approximations induced by particular conditions, such as, for
instance, caused by a nonlinear response of a medium to propagation of a
signal whose motion is otherwise governed by linear equations \cite{Chia,
Kell, Tal}.

In view of the ubiquity of nonlinear equations in physics and the singular
importance of the Schr\"{o}dinger equation (SE), it is not completely
untenable to conceive that the linear Schr\"{o}dinger equation is an
approximation to a more fundamental nonlinear equation. No strong evidence
to the contrary has been provided yet. Meanwhile, this hypothesis has
motivated many authors to either explore fundamental premises that the basic
equation of quantum mechanics might, or even ought to be, nonlinear,
sometimes along with more or less general proposals of accomplishing this in
a consistent manner, or to invent arguments against this case on a
fundamental level. What has indeed been observed is that in some
circumstances this equation naturally acquires a nonlinear extension,
typically in a description of phenomena involving many particles, to name
one but exquisitely important an instance, the Bose-Einstein condensate. In
this case, the leading correction in the mean-field approach to this system
results in the cubic nonlinear Schr\"{o}dinger equation which in the context
under discussion is often referred to as the Gross-Pitaevskii equation \cite
{Gross1, Gross2, Pita}. Therefore, even if the Schr\"{o}dinger equation may
never be found truly nonlinear, the quantum mechanical picture of reality
can equally well be found incomplete without nonlinearities aimed to
describe particular physical effects phenomenologically. As a result of
these two approaches, fundamental and phenomenological, a large and growing
literature originating from different philosphies has been devoted to
nonlinear modifications of quantum mechanics, and, in particular,
modifications of its fundamental equation.

A consistent theoretical framework for nonlinear quantum mechanics (NLQM)
was given by Mielnik \cite{Mie1} who also pointed out \cite{Mie2} that since
there exist many different probability models, they may be related to some
nonlinear variants of quantum mechanics. Even though nonlinear extensions of
the Schr\"{o}dinger equation are usually assumed to depart from its linear
counterpart in some reasonably small way, certain qualitatively new changes
can occur. This is, for instance, demonstrated by the mobility phenomenon 
\cite{Mie3, Mie4} reflecting the fact that the scalar product is not
conserved for quantum states whose evolution is governed by a nonlinear
equation. Two basic and physically important properties of the standard
Schr\"{o}dinger equation are, as a rule, sacrificed in NLQM. One is the
linear superposition principle that can never\footnote{%
In some rare cases \cite{Pusz2, Pusz3} it can be partially retained, though.}
be maintained while the other is the separability of composite systems that
can be preserved in some nonlinear modifications of this equation. It is the
latter property that has acquired a privileged status for it can
discriminate between more and less physically tenable modifications. Indeed,
the lack of separability leads to rather unacceptable physical consequences:
even in the absence of any interactions the motion of one wave packet can
affect the behavior of the other one in a composite system consisting of
these two packets, which clearly violates causality. In its original form 
\cite{Bial}, this condition required that a physically acceptable
nonlinearity should allow two separated, noninteracting and noncorrelated
subsystems to evolve independently of each other. The essential element of
this postulate is that the subsystems are uncorrelated, meaning that the
total wave function is the product of composite wave functions. Equations
that have such a separability property have come to be called weakly
separable \cite{Czach3}. As shown in \cite{Gold}, this property naturally
occurs in a class of nonlinear Schr\"{o}dinger equations characterized by a
certain homogeneity property. The equations of this class which include
those proposed by Kostin \cite{Kos}, Bia\l ynicki-Birula and Mycielski \cite
{Bial}, and Doebner and Goldin \cite{Doeb1, Doeb2} are related via
generalized nonlinear gauge transformations \cite{Doeb3}. Nevertheless, it
was noted \cite{Wein1, Gis1, Gis2, Czach1} that weakly separable equations
may still violate causality if the initial state of a composite system is
entangled.

On the other hand, it was found \cite{Pol, Jor, Czach2} that there exists a
class of deterministic, i.e. non-stochastic, nonlinear generalizations of
quantum mechanics in which the mentioned causality problems can be avoided
for both pure-entangled and general-mixed states if an appropriate
multi-particle extension is used. Such modifications can be called strongly
separable \cite{Czach3}. Based on this observation, a novel general approach
to the issue of separability has recently been put forward by Czachor \cite
{Czach3}. This approach chooses as its starting point the nonlinear von
Neumann equation for the density matrices \cite{Czach4} and proceeds from
there to the $n$-particle extension. In a sense, one can call this approach
``effective'' as opposed to the ``fundamentalist'' one. The latter approach
does not resort to reformulating in terms of density matrices various
proposed in the literature nonlinear modifications of the Schr\"{o}dinger
equation that, similarly as the linear equation itself, are postulated to be
obeyed by pure states. The problem of nonlocality is confounded since these
approaches, even though they belong to the same strong separability
framework, yield different results. For instance, the Bia\l ynicki-Birula
and Mycielski modification that is weakly separable and strongly in the
effective approach turns out to be essentially nonlocal in the
fundamentalist approach for a general nonfactorizable two-particle wave
function as recently demonstrated by L\"{u}cke \cite{Luc1}. The same
applies to the Doebner-Goldin modification that being both weakly and
strongly separable in the effective approach fails to maintain the
separability for a more general nonfactorizable wave function describing a
system of two particles if the fundamentalist approach is used. Only in some
special case, that corresponds to the linearizable Doebner-Goldin equation,
are the particles separable in the latter approach \cite{Luc2}.

The work of Weinberg \cite{Wein1, Wein2}, that proposed a relatively general
framework for possible nonlinear modifications of quantum mechanics, has
renewed and greatly stimulated the interest in various aspects of nonlinear
quantum mechanics and, in particular, in modifications of the fundamental
equation of this theory. It is in Weinberg's framework that the homogeneity
property, adopted from the linear Schr\"{o}dinger equation, becomes elevated
to one of the fundamental assumptions of nonlinear generalization of quantum
mechanics. Although already present in some earlier modifications of this
equation \cite{Ros, Ban, Kib, Smo}, the authors of the proposals following
Weinberg's \cite{Cas, Sab, Doeb1, Doeb2, Aub, Not, Pusz1, Pusz2, Pusz3} have
also considered this property an essential feature of their modified
Schr\"{o}dinger equation. The main general motivation to retain this feature
in the generalized nonlinear framework is to ensure that the departure from
the linear structure of quantum mechanics is not as dramatic as to render
the modified scheme loose all relevant properties of this theory. The basic
physical consequence of preserving this property in a nonlinear version of
the Schr\"{o}dinger equation is that, similarly as in linear theory, any two
wave functions $\Psi $ and $\lambda \Psi $ represent the same physical
state. Weinberg also noted that the homogeneity is sufficient to guarantee
the weak separability of composite systems. However, even if this is true in
his modification and in some others that possess the property in question 
\cite{Gold, Ban, Aub, Pusz2, Pusz3}, it cannot be generically extended as
shown by some example of homogeneous yet nonseparable nonlinear
Schr\"{o}dinger equation \cite{Pusz1}. It has also been demonstrated \cite
{Czach3} that the homogeneity is not in the least necessary for the strong
separability of nonlinear variants of this equation in the efective
approach. Therefore, it seems that, in principle, there is no compelling
reason to require it in the general nonlinear theory. In the light of these
arguments, it is not completely unfounded to conceive that the property in
question is probably an accidental feature of the discussed equation and
thus disposing of it in a nonlinear scheme may not necessarily cause a
tremendous departure from the linear structure of the theory. It is the main
purpose of this paper to demonstrate that this is not the case as indeed
eliminating this property does entail rather dramatic consequences. As we
will see, compromising homogeneity leads to an ambiguity in the expression
for the energy of a quantum-mechanical system that in the case of the linear
Schr\"{o}dinger equation is uniquely determined both as the expectation
value of a Hamiltonian that serves as the generator of evolution and in the
field-theoretical Lagrangian framework as a constant of motion for systems
whose Lagrangian does not depend explicitly on time. In nonlinear theory,
the discussed property is in general sacrificed, meaning that these two
objects represent different quantities. It is the main purpose of this paper
to show that there exists a class of nonlinear modifications of the
fundamental equation of quantum theory that preserve the property in
question and to spell out the conditions under which this occurs.

To this end, in the next section we present the ambiguity in its general
form and point out that there exist a well-defined class of nonlinear
modifications of the Schr\"{o}dinger equation in which it does not arise.
The examples of modifications affected by the problem in question are
discussed in the section that follows as are the examples of the
modifications already put forward in the literature in which the ambiguity
is absent. Our findings are summarized in the conclusions. The appendix
contains a more detailed presentation of how we arrived at the class of
ambiguity-free modifications.

\section{Ambiguity}

In what follows, we will consider nonlinear modifications of the
Schr\"{o}dinger equation that possess a Lagrangian formulation.
Consequently, as just noted, as long as the Lagrangian is not explicitly
time-dependent (that is, for instance, the potentials in which
quantum-mechanical systems evolve are time-independent), the energy of such
systems can in principle be a conserved quantity. The quantity in question
is defined as a space integral over the time-time component of the
corresponding canonical energy-momentum tensor. This energy functional, that
we choose to call field-theoretical, can then be contrasted with the energy
functional which we term quantum-mechanical for it is defined as an
expectation value of the Hamiltonian operator $H$. In NLQM, as we will see,
once the Lagrangian for a modification is defined, one finds this operator
from its equations of motion. In linear quantum mechanics these two
definitions of energy coincide.

We will deal only with wave functions that are square-integrable and thus
normalizable in the norm naturally induced by the scalar product 
\begin{equation}
<\Psi _{1}|\Psi _{2}>\equiv \int d^{3}x\Psi _{1}^{*}(\vec{x})\Psi _{2}(\vec{x%
}).  \label{1}
\end{equation}
Let us denote the total Lagrangian density for a modified Schr\"{o}dinger
equation by $L(R,S)$ and its Hamiltonian operator by $H(R,S)$, where $R$ and 
$S$ stand for the amplitude and the phase of the wave function, $\Psi =R\exp
(iS)$, correspondingly. Now, $L(R,S)=L_{SE}(R,S)-L_{NL}(R,S)$ and $%
H(R,S)=H_{SE}(R,S)+H_{NL}(R,S)$, where subscripts $SE$ and $NL$ denote the
parts leading to or corresponding to a purely linear and a nonlinear
``correction'' of these quantities, respectively. We will use this
hydrodynamic representation \cite{Mad} throughout the rest of the paper. We
assume that the Lagrangian density does not contain differential position
operators of the order higher than second and time derivative terms beyond
those that appear in its linear part. The former assumption is made solely
for the sake of simplifying our considerations. We also assume that the
Lagrangian is a real local scalar functional, bilinear in $\Psi $ in the
sense that it can be presented in the form $F(R,S)R^{2}$, where $F(R,S)$ is
a certain functional of $R$ and $S$, their derivatives, and some external
potential $V$. The Hamiltonian operator can in principle be complex. As a
matter of fact, this is the case even for the Hamiltonian of the ordinary
Schr\"{o}dinger equation. Indeed, 
\begin{equation}
H_{SE}\Psi =\left[ -\frac{\hbar ^{2}\Delta R}{2mR}+\frac{\hbar ^{2}\left( 
\vec{\nabla}S\right) ^{2}}{2m}+\frac{i\hbar ^{2}}{2mR^{2}}\vec{\nabla}\cdot
(R^{2}\vec{\nabla}S)+V\right] \Psi =H_{SE}(R,S)\Psi .  \label{2}
\end{equation}
Usually, unlike in the linear case, the Hamiltonian operator of a modified
equation is not Hermitian. It is sometimes required \cite{Gold} that the
operator in question be norm-Hermitian, i.e., 
\begin{equation}
\langle \Psi |H|\Psi \rangle \equiv \int d^{3}x\Psi ^{*}\left( H\Psi \right)
=\int d^{3}x\left( H\Psi \right) ^{*}\Psi ,  \label{3}
\end{equation}
for any normalizable function $\Psi $ in its domain. This ensures that the
quantum-mechanical energy of a system, defined as the expectation value of
its Hamiltonian $H$ is real. As we will see, in some cases even real
Lagrangian densities that entail a real field-theoretical energy lead to
Hamiltonian operators that are not norm-Hermitian. Therefore, for the sake
of the generality of our considerations, we do not insist that Hamiltonian
operators discussed here possess this property. What we can observe on this
example is that the Hamiltonian operator of a nonlinear modification of the
Schr\"{o}dinger equation is norm-Hermitian if 
\begin{equation}
Im H_{NL}(R,S)=\frac{const}{R^{2}}\vec{\nabla}\cdot (f(R,S)),  \label{4}
\end{equation}
where $f(R,S)$ is a certain functional of $R$ and $S$ that vanishes on the
boundary. It should be pointed out that the Hamiltonian $H_{SE}$ in its
hydrodynamic nonlinear form (2) is not Hermitian, but, in fact, only
norm-Hermitian. This an artifact of the nonlinear Madelung representation.

Knowing the Lagrangian $L(R,S)$ for a modification one derives the equations
of motion for it in the hydrodynamic formulation by varying the Lagrangian
with respect to $R$, $S$, and concomitant derivatives. Since for the linear
Schr\"{o}dinger equation 
\begin{equation}
-L_{SE}(R,S)=\hbar R^{2}\frac{\partial S}{\partial t}+\frac{\hbar ^{2}}{2m}%
\left[ \left( \vec{\nabla}R\right) ^{2}+R^{2}\left( \vec{\nabla}S\right)
^{2}\right] +VR^{2},  \label{5}
\end{equation}
these equations have the form 
\begin{equation}
\hbar \frac{\partial R^{2}}{\partial t}+\frac{\hbar ^{2}}{m}\vec{\nabla}%
\cdot (R^{2}\vec{\nabla}S)+\frac{\delta L_{NL}}{\delta S}=0,  \label{6}
\end{equation}
\begin{equation}
\frac{\hbar ^{2}}{m}\Delta R-\frac{\hbar ^{2}}{m}R\left( \vec{\nabla}%
S\right) ^{2}-2\hbar R\frac{\partial S}{\partial t}-2VR-\frac{\delta L_{NL}}{%
\delta R}=0,  \label{7}
\end{equation}
where $\delta L_{NL}/\delta S$ and $\delta L_{NL}/\delta R$ are Lagrangian
derivatives with respect to $S$ and $R$, correspondingly. If we write down
the modified Schr\"{o}dinger equation in the form resembling that of the
linear equation, 
\begin{equation}
i\hbar \frac{\partial \Psi }{\partial t}=H\Psi ,  \label{8}
\end{equation}
the imaginary part of it will lead to the continuity equation (6) and the
real part to equation (7). Therefore, 
\begin{equation}
H(R,S)=-\frac{\hbar ^{2}\Delta R}{2mR}+\frac{\hbar ^{2}\left( \vec{\nabla}%
S\right) ^{2}}{2m}+V+\frac{1}{2R}\frac{\delta L_{NL}}{\delta R}+\frac{i\hbar
^{2}}{2mR^{2}}\vec{\nabla}\cdot (R^{2}\vec{\nabla}S)+\frac{i}{2R^{2}}\frac{%
\delta L_{NL}}{\delta S},  \label{9}
\end{equation}
from which we recognize that 
\begin{equation}
H_{NL}(R,S)=\frac{1}{R}\frac{\delta L_{NL}}{\delta R}+\frac{i}{2R^{2}}\frac{%
\delta L_{NL}}{\delta S}.  \label{10}
\end{equation}
The quantum-mechanical energy functional is 
\begin{equation}
E_{QM}=\langle \Psi |H|\Psi \rangle =\int d^{3}x\left\{ \frac{\hbar ^{2}}{2m}%
\left[ R^{2}\left( \vec{\nabla}S\right) ^{2}+\left( \vec{\nabla}R\right)
^{2}\right] +VR^{2}+\frac{R}{2}\frac{\delta L_{NL}}{\delta R}+\frac{i}{2}%
\frac{\delta L_{NL}}{\delta S}\right\} ,  \label{11}
\end{equation}
where the term containing $\vec{\nabla}\cdot (R^{2}\vec{\nabla}S)$ has not
been included. Since it is a total derivative, when integrated over the
entire space it produces zero owing to vanishing of $R^{2}$ on the boundary
in the infinity. As already noted, this functional can be complex for the
Hamiltonian operator is not necessarily norm-Hermitian.

The field-theoretical energy functional for the Lagrangian density that
depends on $R$, $S$, and its first and second order derivatives stems from
the following canonical energy-momentum tensor 
\begin{equation}
T_{\nu }^{\mu }=\sum_{i}\left[ \frac{\delta L}{\delta \partial _{\mu
}\varphi _{i}}\partial _{\nu }\varphi _{i}+\frac{\delta L}{\delta \partial
_{\mu }\partial _{\alpha }\varphi _{i}}\partial _{\nu }\partial _{\alpha
}\varphi _{i}-\partial _{\alpha }\left( \frac{\delta L}{\delta \partial
_{\mu }\partial _{\alpha }\varphi _{i}}\partial _{\nu }\varphi _{i}\right)
-\delta _{\nu }^{\mu }L\right] ,  \label{12}
\end{equation}
where $\varphi _{i}=(R,S)$, $i=1,2$. Defined as the space integral over the
time-time component of this tensor, it reads 
\begin{equation}
E_{FT}=\int d^{3}x\left\{ \frac{\hbar ^{2}}{2m}\left[ R^{2}\left( \vec{\nabla%
}S\right) ^{2}+\left( \vec{\nabla}R\right) ^{2}\right]
+VR^{2}+L_{NL}\right\} +C,  \label{13}
\end{equation}
where $C$ is an arbitrary constant that can be put zero and which results
from the integration of the conservation law 
\begin{equation}
\partial _{\mu }T_{\nu }^{\mu }=0.  \label{14}
\end{equation}
We see that in general $E_{QM}\neq E_{FT}$. In particular, this holds true
for the stationary states of a modified Schr\"{o}dinger equation which are
determined by the equation $\partial R^{2}/\partial t=0$. As long as $V\neq
V(t)$, this equation together with (6-7) implies that, similarly as in the
linear case, the stationary states have the form $\Psi =R\exp (-i\hbar
\omega +\sigma (\vec{x}))$, where $\omega $ is the frequency. The energy of
such states, $E_{\omega }$, is by the virtue of the continuity equation real
and equal $E_{QM}=\hbar \omega $, but even in such circumstances $E_{QM}$ is
not always equal to $E_{FT}$.

In its most general form suitable for our purposes, $L_{NL}(R,S)$ can be
represented by $G(R,S)R^{2}$, where $G(R,S)$ is a functional of $R$, $S$,
and their first and second order derivatives. \footnote{%
The expressions like $\left( \vec{\nabla}S\times \vec{\nabla}S\right) ^{2}$, 
$\left( \vec{\nabla}S\times \vec{\nabla}R\right) ^{2}/R^{2}$ or $\left( \vec{%
\nabla}R\times \vec{\nabla}R\right) ^{2}/R^{4}$ are excluded from our
considerations. In general, they would entail the ambiguity.} It can be
shown that only if $G(R,S)=$ $G_{h}(R,S)$, where $G_{h}$ is homogeneous of
degree zero in $\Psi $ and such that 
\begin{equation}
G_{h}(R,S)=\left[ b_{0}+b_{1}\left( \frac{\vec{\nabla}R}{R}\right)
^{2}\right] p(S),  \label{15}
\end{equation}
where 
\begin{equation}
p(S)=a_{1}\left( \vec{\nabla}S\right) ^{2n_{1}}+a_{2}\left( \Delta S\right)
^{n_{2}}+a_{3}\left( \vec{\nabla}S\right) ^{2n_{3}}\left( \Delta S\right)
^{n_{4}},  \label{16}
\end{equation}
and $a_{i}$, $b_{i}$ are real constants and $n_{i}$ non-negative integers,
the ambiguity in question disappears, i.e., 
\begin{equation}
E_{QM}=E_{FT}=\int d^{3}x\left\{ \frac{\hbar ^{2}}{2m}\left[ R^{2}\left( 
\vec{\nabla}S\right) ^{2}+\left( \vec{\nabla}R\right) ^{2}\right]
+VR^{2}+G_{h}(R,S)R^{2}\right\} ,  \label{17}
\end{equation}
with $C=0$. We will denote the Lagrangian of this property by $L_{h}(R,S)$.
The discussed homogeneity of $G_{h}(R,S)$ implies in particular that $G_{h}$
cannot be polynomial in $S$, but needs to depend on $S$ through at least
first order derivatives. Indeed, since $S=\frac{i}{2}\ln \left( \Psi /\Psi
^{*}\right) $, $S(\lambda \Psi ,\left( \lambda \Psi \right) ^{*})\neq S(\Psi
,\Psi ^{*}\dot{)}$ and one can easily show that if $G$ is proportional to $%
S^{k}$, $E_{QM}\neq E_{FT}$. This demonstration and the proof that $%
G_{h}(R,S)$ has to be of the form (15-16) are relegated to the appendix.
This constitutes the main result of the present paper.

\section{Examples}

In what follows, we will adopt the convention $\hbar =1$. To begin with, let
us consider a simple nonhomogenous model Lagrangian density that in its
inhomogenous part employes terms similar to those in the Lagrangian density
for the linear Schr\"{o}dinger equation. The nonlinear part of the
Lagrangian density for the modification is 
\begin{equation}
L_{NL}=a\left( \vec{\nabla}S\right) ^{2}+b\left( \frac{\vec{\nabla}R}{R}%
\right) ^{2},  \label{18}
\end{equation}
where $a$ and $b$ are certain dimensional constants. The equations of motion
read 
\begin{equation}
\frac{\partial R^{2}}{\partial t}+\vec{\nabla}\cdot \left( R^{2}\vec{\nabla}%
S\right) +2a\Delta S=0,  \label{19}
\end{equation}
\begin{equation}
\frac{1}{m}\Delta R+2b\left[ \frac{1}{R}\left( \frac{\vec{\nabla}R}{R}%
\right) ^{2}+\vec{\nabla}\cdot \left( \frac{\vec{\nabla}R}{R^{2}}\right)
\right] -\frac{1}{m}R\left( \vec{\nabla}S\right) ^{2}-2R\frac{\partial S}{%
\partial t}-2VR=0.  \label{20}
\end{equation}
From these the nonlinear part of the Hamiltonian is identified with 
\begin{equation}
H_{NL}=-i\frac{a\Delta S}{R^{2}}+b\frac{\Delta \ln R^{2}}{R^{2}}.  \label{21}
\end{equation}
The quantum-mechanical energy of a system described by $R$ and $S$ is then 
\begin{equation}
E_{QM}=\int d^{3}x\left\{ \frac{1}{2m}\left[ \left( \vec{\nabla}R\right)
^{2}+R^{2}\left( \vec{\nabla}S\right) ^{2}\right] +VR^{2}+b\Delta \ln
R^{2}-ia\Delta S\right\} .  \label{22}
\end{equation}
However, one finds that the field-theoretical energy functional for this
case is 
\begin{equation}
E_{FT}=\int d^{3}x\left\{ \frac{1}{2m}\left[ \left( \vec{\nabla}R\right)
^{2}+R^{2}\left( \vec{\nabla}S\right) ^{2}\right] +VR^{2}+a\left( \vec{\nabla%
}S\right) ^{2}+b\left( \frac{\vec{\nabla}R}{R}\right) ^{2}\right\} .
\label{23}
\end{equation}
To convince ourselves that these two expressions for the energy give
different results let us set\footnote{%
This model of NLSE should not be construed as anything else than a toy model
intended solely for the purpose of this exposition. To make it less toyish
one should put $b=0$ anyway so as to allow a greater class of physically
acceptable solutions which for an arbitrary $b\neq 0$ would be excluded by
the condition of finite energy.} $b=0$ and consider the case of a
one-dimensional coherent state wave packet. Such a packet requires the
potential of a simple harmonic oscillator, $V=m\omega ^{2}x^{2}/2$. Its
amplitude and phase are given by ($x_{0}=1/\sqrt{m\omega }$.) 
\begin{equation}
R_{coh}^{2}=\frac{1}{\sqrt{\pi }x_{0}}\exp \left[ {-\frac{\left( x-x_{0}%
\sqrt{2}\cos (\omega t-\delta )\right) ^{2}}{{x_{0}}^{2}}}\right]  \label{24}
\end{equation}
and 
\begin{equation}
S_{coh}=-\left( \frac{\omega t}{2}-\frac{|\alpha |^{2}}{2}\sin 2\left(
\omega t-\delta \right) +\frac{\sqrt{2}|\alpha |x}{x_{0}}\sin \left( \omega
t-\delta \right) \right) ,  \label{25}
\end{equation}
correspondingly, where $\alpha $ and $\delta $ are arbitrary numbers,
complex and real, respectively. The coherent state is a solution to our
equations of motion as its phase satisfies $\Delta S_{coh}=0$. Now, since $%
\vec{\nabla}S_{coh}$ is not identically zero, the expressions in question
produce completely different values for the energy of this state. In fact, $%
E_{FT}$ is infinite for most of the time! One might think that this
modification is too contrived. However, a similar situation occurs in the
Staruszkiewicz modification of the Schr\"{o}dinger equation \cite{Star}
which can be derived from the Lagrangian 
\begin{equation}
-L_{SM}=R^{2}\frac{\partial S}{\partial t}+\frac{1}{2m}\left[ \left( \vec{%
\nabla}R\right) ^{2}+R^{2}\left( \vec{\nabla}S\right) ^{2}\right] +R^{2}V+%
\frac{c}{2}\left( \Delta S\right) ^{2}  \label{26}
\end{equation}
leading to the equations 
\begin{equation}
\frac{\partial R^{2}}{\partial t}+\frac{1}{m}\vec{\nabla}\cdot \left( R^{2}%
\vec{\nabla}S\right) -c\Delta \Delta S=0,  \label{27}
\end{equation}
\begin{equation}
\frac{1}{m}\Delta R-R\left( \vec{\nabla}S\right) ^{2}-2R\frac{\partial S}{%
\partial t}-2VR=0.  \label{28}
\end{equation}
The field-theoretical form of the energy functional for this modification, 
\begin{equation}
E_{FT}=\int \,d^{3}x\,\left\{ \frac{1}{2m}\left[ \left( \vec{\nabla}R\right)
^{2}+R^{2}\left( \vec{\nabla}S\right) ^{2}\right] +\frac{c}{2}\left( \Delta
S\right) ^{2}+VR^{2}\right\} ,  \label{29}
\end{equation}
is again different from its quantum-mechanical counterpart, 
\begin{equation}
E_{QM}=\int d^{3}x\left\{ \frac{1}{2m}\left[ \left( \vec{\nabla}R\right)
^{2}+R^{2}\left( \vec{\nabla}S\right) ^{2}\right] +VR^{2}+\frac{ic}{2}\Delta
\Delta S\right\} ,  \label{30}
\end{equation}
which can be easily obtained from the Hamiltonian for this modification 
\begin{equation}
H_{SM}=H_{SE}+\frac{ic\Delta \Delta S}{2R^{2}}.  \label{31}
\end{equation}
Now, even if these two energy functionals are equal for the coherent state
they do drastically differ for ordinary Gaussian wave packets for which $%
\Delta S=g(t)$. Obviously, these wave packets are solutions to the equations
of motion of the modification.

The energy functionals $E_{QM}$ discussed so far contain imaginary
components. Since the energy is supposed to be a real quantity one might
want to require that these parts do not contribute to the total energy,
which imposes a constraint on physically acceptable states allowed by these
particular nonlinear models of the Schr\"{o}dinger equation. That these
constraints are not necessarily very restrictive or physically unjustified
can be seen from the Staruszkiewicz modification for which $E_{FT}$ is
infinite for Gaussian wave packets which therefore should be excluded if
this energy definition were employed and which are perfectly fine on the
energetic grounds if one uses the quantum-mechanical definition of energy.
Moreover, the continuity equation implies vanishing of the imaginary terms
for a large class of physically interesting situations. The condition of
real energy in the cases presented (if one assumes a less restricting case
of $b=0$ in the toy model described by (19) and (20)) is equivalent to the
selection of observables in NLQM, or, more precisely, of their domain, for
which contributions of nonlinear parts vanish on normalized states. This in
itself does not resolve the problem of ambiguity as even if $E_{QM}$ is now
real it is still in general different from $E_{FT}$. What it does though is
to illustrate the fact that the properties of observables depend on the
space of states on which they are defined. This is so in particular in
linear QM, where, for instance, the self-adjointness of an operator depends
on its domain. It is in line with this observation that one can attempt to
reconcile $E_{QM}$ with $E_{FT}$ by choosing a domain in which they are
equal for each function in this domain. Nevertheless such an alternative
would be overly restrictive in some cases, to the extent that it would
result in a trivial and physically unsatisfactory domain. The simplest
example of the modification of this kind is provided by the well-known
nonlinear cubic Schr\"{o}dinger equation for which $L_{NL}=R^{4}$. One could
also completely reject the statistical definition of energy in quantum
theory, $E_{QM}$, but this would constitute a very dramatic departure from
the conceptual structure of this theory. In fact, by identifying the class
of modifications for which $E_{QM}=E_{FT}$, we demonstrated that this is not
necessary.

We have limited our considerations to local Lagrangian densities, but it can
be shown that the ambiguity in question occurs also in modifications that
stem from nonlocal nonhomogeneous Lagrangians. For some particular form of
the nonlinear part of the Lagrangian density, 
\begin{equation}
L_{NL}^{^{\prime }}=G(\rho )\rho =\int_{0}^{\rho }d\rho ^{\prime }F(\rho
^{\prime }),  \label{32}
\end{equation}
this has been demonstrated in \cite{Bial}. This Lagrangian leads to the
Hamiltonian $H_{NL}^{^{\prime }}=F(\rho )$, where $F$ is a functional of $%
\rho =R^{2}$ and it is assumed that it contains no derivatives. One has 
\begin{equation}
E_{QM}-E_{FT}=<\Psi |F|\Psi >-<\Psi |G|\Psi >-C=D-C.  \label{33}
\end{equation}
The discussed class of nonhomogeneous modifications includes as a special
case the Bia{\l }ynicki-Birula and Mycielski modification \cite{Bial}
characterized by $F(\rho )=c_{1}\ln c_{2}\rho $, where $c_{1}$ and $c_{2}$
are constants. This modification is an exception to the rule as here $%
D=c_{1} $ for wave functions normalized to unity and so by choosing $c_{1}=C$
one causes the ambiguity to disappear. The Bia\l ynicki-Birula and Mycielski
modification can in fact be thought of as a homogeneous modification modulo
the phase gauge transformation $S\rightarrow S-2i\ln |\lambda |$ which when
exercised along with the homogeneity transformation $\Psi \rightarrow
\lambda \Psi $ renders the NLSE of the modification homogeneous. It is also
this gauge transformation that removes the difference between $E_{QM}$ and $%
E_{FT}$. Moreover, as shown in \cite{Bial}, if $E_{QM}=E_{FT}$ for all
stationary states of $H=H_{SE}+H_{NL}^{^{\prime }}$ then the nonlinearity
must be logarithmic as in the modification under consideration, which proves
the generic nature of the ambiguity in the class of nonlinear modifications
derivable from the nonlocal Lagrangian (32).

Let us now discuss the examples of modifications known in the literature
that derive from local Lagrangian densities in which the ambiguity problem
does not arise. The first of them constitutes a restricted version of the
Doebner-Goldin modification \cite{Doeb1, Doeb2} for which the Lagrangian
density was found in \cite{Pusz1}. The density in question reads 
\begin{equation}
L_{DG}^{r}(R,S)=L_{SE}+c_{1}R^{2}\Delta S.  \label{34}
\end{equation}
Its nonlinear part is clearly a special version of $L_{h}(R,S)$ for which $%
b_{1}$ in (15), $a_{1}$ and $a_{3}$ in (16) are zero, and $n_{2}$ in (16) is 
$1$. A more general Lagrangian that encompasses $L_{DG}^{r}$ as a special
case was introduced in \cite{Pusz1}. It is given by 
\begin{equation}
L_{PH}(R,S)=L_{SE}+c_{1}R^{2}\left( \Delta S\right) ^{n}  \label{35}
\end{equation}
and corresponds to special but fairly general a Galilean invariant version
of our $L_{h}(R,S)$ with $b_{1}=a_{1}=a_{3}=0$ in (15-16) and an arbitrary $%
n_{2}=n$ in (16). The particular case of $n=2$ was treated in more detail in 
\cite{Pusz1, Pusz4}. It should be noted that a generalization of $p(S)$ of
(16) is still possible; the terms $\left( \Delta ^{m}S\right) ^{n}$, where $%
m>1$, and their products with $\left( \vec{\nabla}S\right) ^{2k}$ can be
added to $p(S)$ without violating the equivalence of $E_{QM}$ and $E_{FT}$.
The latter case represents the most general scheme proposed in \cite{Pusz1}.

\section{Conclusions}

We have noted and discussed the ambiguity in the definition of the energy
functional in NLQM. As opposed to linear theory, the energy functional
defined as the expectation value of the Hamiltonian operator is not equal to
the conserved quantity derived within the Lagrangian field-theoretical
framework that one identifies with the energy of a system. It has been shown
that inspite of the generic nature of this ambiguity, there exists a class
of nonlinear modifications of the Schr\"{o}dinger equation derivable from
local Lagrangian densities in which the ambiguity in question does not
arise. This is tantamount to the definition of a new class of such
modifications uniquely characterized by the property in question. The
modifications of this class must be homogeneous in the wave function and, as
a rule, they involve spatial derivatives of its phase in their NLSE. It is
the presence of the phase functionals in the Lagrangian densities that is
the most salient feature of this class. Paradoxically enough, the phase
modifications of the Schr\"{o}dinger equation have not been particularly
pursued in the literature. Only two examples of the class under
consideration have been proposed so far \cite{Doeb2, Pusz1}, although their
exceptionality in the sense discussed have not been fully realized.

We see therefore that even if the homogeneity of NLSE is neither necessary
for its strong \cite{Czach3} nor sufficient for its weak separability \cite
{Pusz1}, preserving this property in the NLSE certainly adds to the
congruency of the formulation of NLQM and warrants that one of highly
desirable features of linear theory, the non-ambiguous definition of energy,
does not have to be abandoned, but can be retained this special scheme of
modifications. It also makes this type of modifications more similar to the
linear SE, and saves us from resorting to more radical means of resolving
this problem as, for instance, rejecting the statistical interpretation of
quantum theory and, consequently, $E_{QM}$ as a viable definition of energy.
As found in \cite{Pusz1} for some Galilean invariant subclass of this
scheme, the modifications in question do not have the standard Ehrenfest
limit, which suggests that they are not linearizable and thus can describe
some new physics that the linear theory is unable to capture. Further
studies are necessary to understand all the physical implications of this
unique property that defines the discussed class of modifications.

Let us note that despite the fact that the nonlinear Hamiltonians we
considered are manifestly non-Hermitian, their expectation values are real
numbers. Although relatively little known, this also happens to be true for
linear operators \cite{Grub}, and, quite recently, it has been observed for
some surprisingly simple complex-valued one-dimensional potentials \cite
{Ben1}. The Hermiticity of Hamiltonians is sometimes also purposedly
sacrificed in linear QM in order to simplify description of complex
phenomena involving many degrees of freedom not all of which can be taken
into account. For instance, it is well known that to describe absorption in
scattering processes, such as elastic scattering in nuclear physics, one can
implement complex or ``optical'' potentials. Such potentials can also be
used to describe decoherence \cite{Mens}. It should also be noted that not
all equations of physical interest can be derived from local Lagrangian
densities; the best case in point is provided by the celebrated
Navier-Stokes equations for which no Lagrangian exists \cite{Fin}.

In the course of this work, we became aware that the discussed energy
ambiguity was observed by other authors as well. We have already alluded to
this when discussing the modifications that stem from nonlocal Lagrangian
densities. The special case of such modifications was discussed by Bia\l
ynicki-Birula and Mycielski \cite{Bial} mainly as a way to emphasize the
uniqueness of their model. Some more attention this issue received in the
work of Shapiro \cite{Shap}. Both papers considered only real nonlinear
Lagrangian corrections and none of them examined general conditions under
which $E_{QM}=E_{FT}$. Neither did they point out the relevance of the
homogeneity as an important prerequisite for the equivalence of the energy
functionals in question.

\section{Appendix}

We will demonstrate here that the most general form of $G(R,S)$ that
satisfies our conditions and ensures that $E_{QM}=E_{FT}$ is of the form
(15-16). The general nonhomogeneous form of $G(R,S)$ is 
\[
G(R,S)=G_{h}(R,S)\left[
\sum_{k=1}S^{k}+\sum_{l=1}R^{l}+\sum_{m,n=1}S^{m}R^{n}\right] , 
\]
where the homogeneous part can be presented as 
\[
G_{h}(R,S)=p(S)+q(R)+t(R,S), 
\]
with 
\[
p(S)=a_{1}\left( \vec{\nabla}S\right) ^{2n_{1}}+a_{2}\left( \Delta S\right)
^{n_{2}}+a_{3}\left( \vec{\nabla}S\right) ^{2n_{3}}\left( \Delta S\right)
^{n_{4}}, 
\]
\[
q(R)=b_{1}\left( \frac{\vec{\nabla}R}{R}\right) ^{2k_{1}}+b_{2}\left( \frac{%
\Delta R}{R}\right) ^{k_{2}}+b_{3}\left( \frac{\vec{\nabla}R}{R}\right)
^{2k_{3}}\left( \frac{\Delta R}{R}\right) ^{k_{4}}, 
\]
\[
t(R,S)=cp(S)q(R), 
\]
and where $a_{i}$, $b_{i}$, $c$ are some constants. The exponents $m$ and $n$
are in general different from the ones used below in some particular cases.
For the sake of completeness, let us recall that 
\[
E_{QM}=\int d^{3}x\left\{ \frac{\hbar ^{2}}{2m}\left[ R^{2}\left( \vec{\nabla%
}S\right) ^{2}+\left( \vec{\nabla}R\right) ^{2}\right] +VR^{2}+\frac{R}{2}%
\frac{\delta L_{NL}}{\delta R}+\frac{i}{2}\frac{\delta L_{NL}}{\delta S}%
\right\} 
\]
and 
\[
E_{FT}=\int d^{3}x\left\{ \frac{\hbar ^{2}}{2m}\left[ R^{2}\left( \vec{\nabla%
}S\right) ^{2}+\left( \vec{\nabla}R\right) ^{2}\right]
+VR^{2}+L_{NL}\right\} . 
\]
To begin with, let us first discuss the case of $q_{1}(R)=b_{1}\left( \vec{%
\nabla}R/R\right) ^{2m}R^{2}$. One obtains that 
\[
\frac{R}{2}\frac{\delta q_{1}(R)}{\delta R}=\frac{\left( 1-2m\right) b_{1}}{2%
}\left( \frac{\vec{\nabla}R}{R}\right) ^{2m-2}R^{2}\left[ (2-2m)\left( \frac{%
\vec{\nabla}R}{R}\right) ^{2}+2m\left( \frac{\Delta R}{R}\right) \right] 
\]
and consequently, since $\int d^{3}xR\Delta R=$ $-\int d^{3}x\left( \vec{%
\nabla}R\right) ^{2}$, it is only for $m=1$ that $%
E_{QM}[q_{1}(R)]=E_{FT}[q_{1}(R)]$. For $q_{2}(R)=b_{2}\left( \Delta
R/R\right) ^{n}R^{2}$ we derive that 
\begin{eqnarray*}
\frac{R}{2}\frac{\delta q_{2}(R)}{\delta R} &=&\frac{b_{2}R^{2}}{2}\left\{
A_{2}(R)\left( \frac{\Delta R}{R}\right) ^{n}+\right. \\
&&\left. n\left[ B_{2}(R)\left( \frac{\Delta R}{R}\right)
^{n-1}+C_{2}(R)\left( \frac{\Delta R}{R}\right) ^{n-2}+D_{2}(R)\left( \frac{%
\Delta R}{R}\right) ^{n-3}\right] \right\}
\end{eqnarray*}
where 
\[
A_{2}(R)=(2-n)(1+n), 
\]
\[
B_{2}(R)=(2-n)(1-n)\left( \frac{\vec{\nabla}R}{R}\right) ^{2}, 
\]
\[
C_{2}(R)=(n-1)\left[ \frac{\Delta ^{2}R}{R}+2(2-n)\frac{\vec{\nabla}R}{R}%
\cdot \frac{\vec{\nabla}\Delta R}{R}\right] , 
\]
\[
D_{2}(R)=(n-1)(n-2)\left( \frac{\vec{\nabla}\Delta R}{R}\right) ^{2}. 
\]
One can see that under no circumstances $E_{QM}[q_{2}(R)]=E_{FT}[q_{2}(R)]$.
We expect thus that the same happens for $q_{3}(R)=b_{3}\left( \vec{\nabla}%
R/R\right) ^{2m}\left( \Delta R/R\right) ^{n}R^{2}$. This is indeed the case
as 
\begin{eqnarray*}
\frac{R}{2}\frac{\delta q_{3}(R)}{\delta R} &=&\frac{b_{3}}{2}\left( \frac{%
\vec{\nabla}R}{R}\right) ^{2m-2}R^{2}\left\{ A_{3}(R)\left( \frac{\Delta R}{R%
}\right) ^{n}+\right. \\
&&\left. n\left[ B_{3}(R)\left( \frac{\Delta R}{R}\right)
^{n-1}+C_{3}(R)\left( \frac{\Delta R}{R}\right) ^{n-2}+D_{3}(R)\left( \frac{%
\Delta R}{R}\right) ^{n-3}\right] \right\} ,
\end{eqnarray*}
where 
\[
A_{3}(R)=\alpha \left[ 1-2m+n(1+4m)\right] \left( \frac{\vec{\nabla}R}{R}%
\right) ^{2}-2m(1-2m)(n-1)\frac{\Delta R}{R}, 
\]
\[
B_{3}(R)=\alpha (\alpha -1)\left( \frac{\vec{\nabla}R}{R}\right)
^{4}+2m(2n-3)\frac{\vec{\nabla}R}{R}\cdot \frac{\vec{\nabla}\Delta R}{R}, 
\]
\[
C_{3}(R)=(n-1)\left[ \frac{\Delta ^{2}R}{R}+2\alpha \frac{\vec{\nabla}R}{R}%
\cdot \frac{\vec{\nabla}\Delta R}{R}\right] \left( \frac{\vec{\nabla}R}{R}%
\right) ^{2}, 
\]
\[
D_{3}(R)=(n-1)(n-2)\left( \frac{\vec{\nabla}\Delta R}{R}\right) ^{2}\left( 
\frac{\vec{\nabla}R}{R}\right) ^{2}, 
\]
and $\alpha =2-2m-n$. For these reasons it is only $t(R,S)=cp(S)q_{1}(R)$
that can guarantee that $E_{QM}[t(R,S)]=E_{FT}[t(R,S)]$. To see why it is so
and why $p(S)q_{2}(R)$ would not work let us consider an example that also
proves that $p(S)R^{2}$ does not lead to the ambiguity in question. Let us
take 
\[
L_{NL}^{*}=p(S)\left[ R^{2}+b_{1}\left( \vec{\nabla}R\right)
^{2}+b_{2}\left( \Delta R\right) ^{2}\right] =p(S)q^{*}(R). 
\]
Now, 
\[
\frac{\delta L_{NL}^{*}}{\delta S}=-\vec{\nabla}\cdot \left[ \frac{\partial
p(S)}{\partial \vec{\nabla}S}q^{*}(R)\right] +\Delta \left[ \frac{\partial
p(S)}{\partial \Delta S}q^{*}(R)\right] 
\]
which as a total derivative of $R$ and its concomitants does not contribute
to the quantum-mechanical energy functional. We also have 
\[
\frac{R}{2}\frac{\delta L_{NL}^{*}}{\delta R}=p(S)R^{2}-b_{1}R\vec{\nabla}%
\cdot \left( p(S)\vec{\nabla}R\right) +b_{2}R\Delta \left( p(S)\Delta
R\right) 
\]
and since $R\vec{\nabla}\cdot \left( p(S)\vec{\nabla}R\right) =\vec{\nabla}%
\cdot \left( p(S)R\vec{\nabla}R\right) -p(S)\left( \vec{\nabla}R\right) ^{2}$%
, $\int d^{3}xR\vec{\nabla}\cdot \left( p(S)\vec{\nabla}R\right) =-\int
d^{3}xp(S)(\vec{\nabla}R)^{2}$. Because of that, one can have $%
E_{QM}[L_{NL}^{*}]=E_{FT}[L_{NL}^{*}]$, but only if $b_{2}=0$.

We will now consider the general case. It is easy to understand why the
general nonhomogeneous form of $G(R,S)$ would generate the energy ambiguity
even if $G_{h}(R,S)$ is chosen in such a way that it does not cause it
itself. Let us put $m$ and $n$ in the expression for $G(R,S)$ equal zero and
treat first the case $l=0$. One notes that $\delta [G(R,S)R^{2}]/\delta S$
contains terms proportional to $S^{k-1}$ that are not total derivatives. As
a result, $E_{QM}$ is complex as opposed to $E_{FT}$ which is always real.
On the other hand, even if $k=0$, the absence of the ambiguity requires that 
\[
\int d^{3}xR\frac{\delta [G(R,S)R^{2}]}{2\delta R}=\int d^{3}xG(R,S)R^{2}, 
\]
which happens only if $l=0$. We see now that the case of $m\neq 0$ and $%
n\neq 0$ only compounds the problem.

\section*{Acknowledgments}

I would like to thank Professor Pawe{\l } O. Mazur for his critical reading
of an earlier version of the manuscript of this paper, Professor Andrzej
Staruszkiewicz for a stimulating discussion, and Kurt Ko\l tko for his
interest in this work. I am also indebted to Professor Wolfgang L\"{u}cke for a 
correspondence about the issue of separability in nonlinear quantum mechanics. 
A correspondence with Dr. Marek Czachor concerning the paper presented and 
NLQM in general is greatly acknowledged as is a correspondence with Drs. 
Carlos Castro and Ronald Mirman and their interest in this work. This work 
was partially supported by the NSF grant No. 13020 F167 and the ONR grant 
R\&T No. 3124141.

\end{document}